\documentclass[sigconf]{acmart}

\usepackage{booktabs} % For formal tables

 \renewcommand\footnotetextcopyrightpermission[1]{} % removes footnote with conference information in first column
\usepackage{mathtools}

\usepackage{arydshln,leftidx}
\usepackage{amsmath}
\usepackage{amssymb}
\usepackage{kbordermatrix}
\usepackage{graphics}
\usepackage{graphicx}
\usepackage{enumitem}
\usepackage{textcomp}
\usepackage{color}
\usepackage{outlines}

\usepackage{listings}
\definecolor{dkgreen}{rgb}{0,0.6,0}
\definecolor{gray}{rgb}{0.5,0.5,0.5}
\definecolor{mauve}{rgb}{0.58,0,0.82}
\usepackage{amsthm}

\newtheorem{algorithm}{Algorithm}[section]

\usepackage{caption}
\usepackage{setspace}
\usepackage{threeparttable}

\usepackage{changepage}   % for the adjustwidth environment

\usepackage{etoolbox}
\let\bbordermatrix\bordermatrix
\patchcmd{\bbordermatrix}{8.75}{4.75}{}{}
\patchcmd{\bbordermatrix}{\left(}{\left|}{}{}
\patchcmd{\bbordermatrix}{\right)}{\right|}{}{}

\usepackage{blkarray}
\usepackage{multirow}
\usepackage{wrapfig}
\usepackage{tikz}

\usepackage{subfig}
\usetikzlibrary{matrix,fit}

\begin{document}
\title{Post-Quantum Cryptographic Hardware Primitives \vspace{-0.1in}}

\author{Lake Bu, Rashmi Agrawal, Hai Cheng, and Michel A. Kinsy\\
Adaptive and Secure Computing Systems Laboratory\\
Department of Electrical and Computer Engineering, Boston University\\
(bulake, rashmi23, chenghai, mkinsy)@bu.edu\vspace{-0.05in}}

\begin{abstract}
The development and implementation of post-quantum cryptosystems have become a pressing issue in the design 
of secure computing systems, as general quantum computers have become more feasible in the last two years. In this work, we introduce a set of hardware post-quantum cryptographic primitives (PCPs) consisting of four frequently used security components, i.e., public-key cryptosystem (PKC), key exchange (KEX), oblivious transfer (OT), and zero-knowledge proof (ZKP). In addition, we design a high speed polynomial multiplier to accelerate these primitives. These primitives will aid researchers and designers in constructing quantum-proof secure computing systems in the post-quantum era. 
\vspace{-0.15in}
\end{abstract}

\keywords{Post-quantum cryptography, public-key system, key exchange, oblivious transfer, zero-knowledge proof, FPGA-based prototyping.\vspace{-0.1in}}

\maketitle

%%%% Sections 
%\vspace{-0.1in}
\section{Introduction}

In the last three years, we have witnessed a raft of breakthroughs and several key milestones towards the development of general quantum computers. These advances do bring with them critical challenges to classical cryptosystems like RSA (Rivest-Shamir-Adleman), ECC (Elliptic Curve Cryptography), and ElGamal. The strength of these classic algorithms rests on the hardness of integer factorization and discrete logarithm problems, which do not hold under quantum computing approaches. Thus, researchers have been actively investigating new algorithms and designs for cryptosystems for the post-quantum era. Among these techniques, designs based on Ring-learning with errors (Ring-LWE) \cite{LV2010} thus far have proven to be the most promising approach.
Ring-LWE-based cryptosystems have the following advantages (i) their security reduction is a modification of the shortest vector problem (SVP) and closest vector problem (CVP), which are known to be NP-hard, and so far there are no efficient classical or quantum algorithms to solve them; (ii) they can support homomorphic encryption (HE) schemes; (iii) they have much smaller key size comparing with other cryptosystems; (iv) finally, in some cases, they lend themselves to more efficient hardware implementations than their classical competitors.       
In contrast to the extensive literature on the study and software implementation of the Ring-LWE algorithm, there 
has been little work on its efficient hardware implementation. Recently, a handful of works have explored the FPGA implementation of the KEX \cite{OT2017}, and even less the PKC \cite{RS2014}. There is also a general lack of discussion on the design and hardware implementation of other cryptographic primitives such as oblivious transfer (OT) and zero-knowledge proof (ZKP), which play critical roles in many applications such as private machine learning and crypto-currencies using blockchain.\\
\quad Therefore, in this work, we construct a small representative set of reusable, standalone hardware modules of these post-quantum cryptographic primitives (PCPs). They can serve as the fundamental building blocks for a wide range of secure systems. In the work we demonstrate (1) a high speed polynomial multiplier design to aid in the efficient hardware implementation of these primitives, and (2) new algorithms for the OT and ZKP primitives. 

\vspace{-0.05in}
\section{The PCP Hardware Primitives}
\label{sec:algorithms}
\vspace{-0.05in}
\subsection{The Public-Key Cryptosystem (PKC) and Key Exchange (KEX) Primitives} \label{sec: PKC}
The detailed algorithms of the public-key cryptosystem (PKC) and key exchange (KEX) can be found in \cite{LV2010} and \cite{AE2016}, respectively. For brevity, we will only briefly introduce the PKC algorithm, since many of its sub-modules are reused in the KEX primitive. 

\vspace{-0.05in}
\begin{algorithm}
	\normalfont
	Let the ring $R_q$ be $R_q = R/\langle q \rangle = \mathbb{Z}_q[x]/\langle f(x) \rangle$, where $f(x) = x^n + 1$ is an irreducible polynomial with $n$ a power of 2, and $q \equiv 1 \text{ mod } 2n$ is a large prime number. Thus $R_q$ is a ring of integer polynomials modulo both $f(x)$ and $q$, and it has $q^n$ elements. Let $\mathcal{X}$ be a Gaussian distribution of ``small'' errors/noise. If $t = \lfloor \frac{q}{2} \rfloor$, $a, b \in R_q$ and $s, e, r_0, r_1, r_2 \leftarrow \mathcal{X}$, then the public key encryption protocol between Alice and Bob is as follows.
	
	\textbf{Key generation:} 
	Alice picks $s$ and a random $e$ to generate the public key $pk = \{a, b\}$ and the private key $sk = \{s\}$ by:
	\vspace{-0.05in}
	\begin{equation} \label{eq: keygen}
		b = a \cdot s + e 
	\vspace{-0.05in}
	\end{equation}
	
	\textbf{Encryption:} Bob converts his message (plaintext) into a binary vector $m$ of length $n$, and generates the cipher $\{c_0, c_1\}$ as:
	\vspace{-0.05in}
	\begin{equation} \label{eq: enc}
		\begin{cases}
			c_0 & = b \cdot r_0 + r_2 + tm, \\
			c_1 & = a \cdot r_0 + r_1.
		\end{cases}
	%	\vspace{-0.05in}
	\end{equation}

	\textbf{Decryption:} Alice decrypts the cipher by:
	\vspace{-0.05in}
	\begin{equation} \label{eq: dec}
		m = \lceil (c_0 - c_1 \cdot s)/t \rfloor,
	\end{equation}
	where $\lceil \rfloor$ stands for taking the nearest binary integer. 
	
\end{algorithm}

\begin{figure*}[!t]
	\begin{center}
		\includegraphics[width=6.5in]{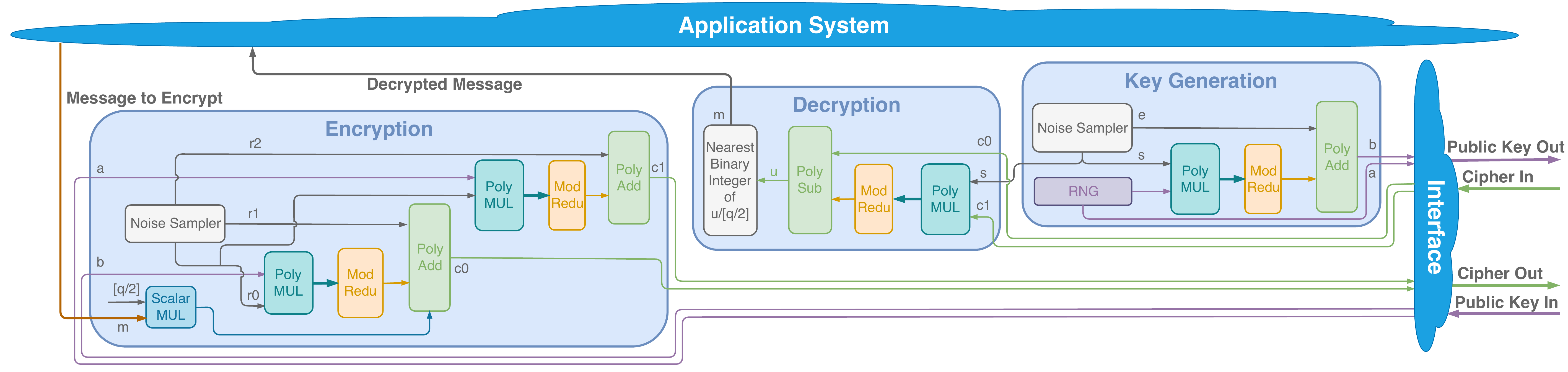} 
	\end{center}
	\vspace{-0.1in}
	\caption{\small The three core building blocks for the primitives: \textit{Key Generation (KeyGen)}, \textit{Encryption (Enc)}, and \textit{Decryption (Dec)}.} 
	\label{fig:pcps}
	\vspace{-0.15in}
\end{figure*}

\vspace{-0.05in}
The basic operations of the algorithms are: polynomial addition, polynomial subtraction, scalar multiplication, scalar division then taking the nearest binary integers, and polynomial multiplication. Most of the operations are component-wise, or can be reduced to conditional assignment. The polynomial multiplication operation has the highest hardware implementation complexity. An efficient multiplication module will substantially improve the hardware implementation efficiency of the entire hardware crypto-primitive suite. Figure \ref{fig:pcps} shows a system architecture using the commonly shared hardware modules. 
%\vspace{-0.05in}

One of the common implementations of the polynomial multiplier, is negative wrapped convolution combined with butterfly number-theoretic transform (NTT, the finite field version of FFT). This approach takes $O(n\text{log}n)$ multiplications and has a time complexity of $O(\text{log}n)$. 
In this work, we are introducing a new and high-speed design of the modular polynomial multiplier, named \textit{Preemptive Adaptive Reduction Multiplier} (PARM). 

It calculates the generalized representation of the product in advance. Thus, given two polynomial multiplicands, their product can be computed as fast as in one step. 
\begin{figure}[!h]
	\vspace{-0.2in}
		\begin{center}
			\includegraphics[width=3.1in]{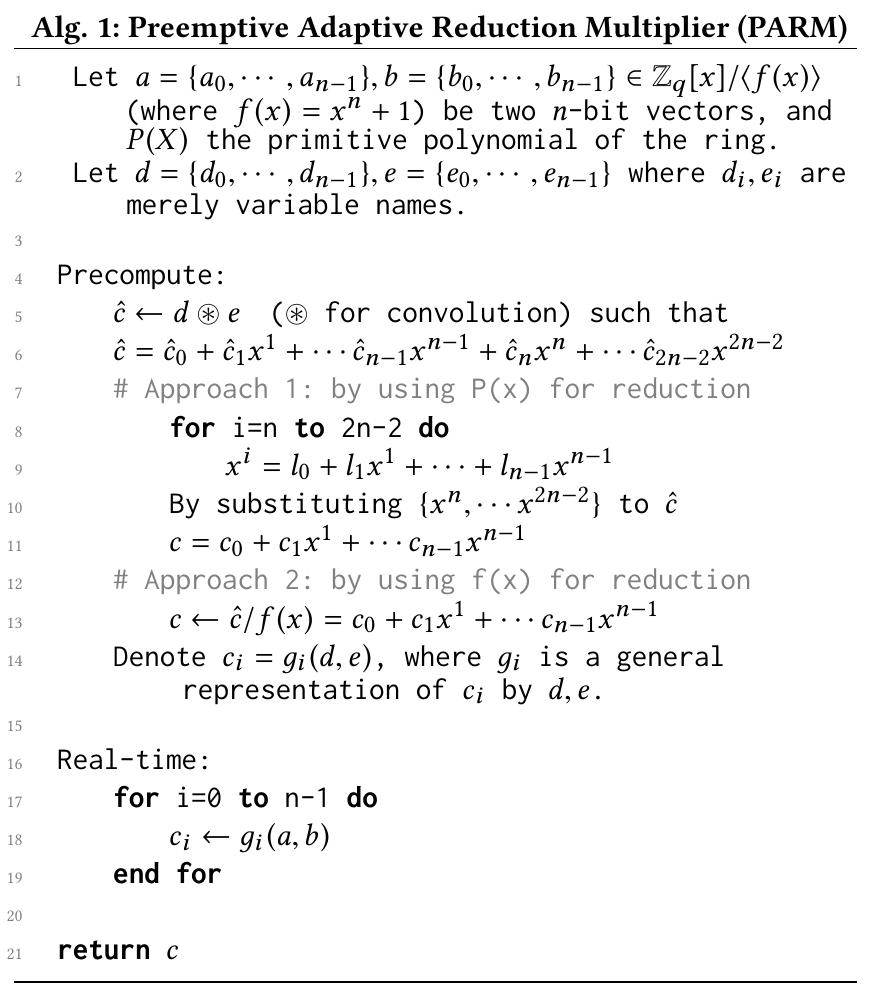} 
		\end{center}
	\vspace{-0.2in}
\end{figure}

As shown in the algorithm 1 outline, the bulk of the work is performed in the ``Precompute'' stage, which is done only once in the lifetime of the multiplier's. The real time computation is just $n$ calculations of $c_i \gets g_i(a, b)$, which can be done fully in parallel (1 cycle) given enough multipliers ($n^2$). 
The resource and time complexities for the PARM algorithm are $O(n^2)$ multiplications and $O(1)$ cycles latency, while the NTT-based complexities are $O(n\text{log}n)$ multiplications and $O(\text{log}n)$ cycles latency. 

\vspace{-0.05in}
\subsection{The Oblivious Transfer (OT) Primitive}
\vspace{-0.05in}
The OT mechanism enables a receiver to choose and receive a certain piece of information out of many pieces from the sender, while remaining oblivious to the other pieces. The sender is also oblivious to the exact piece selected. The OT is a widely used protocol in privacy-preserving computations between two or multiple parties. 
The proposed OT primitive is constructed on the foundation of the PKC primitive. We denote $KeyGen(s, a) = b$ as the \textit{Key Generation} module, $Enc_{pk}(m) = \{c_0, c_1\}$ as the \textit{Encryption} function, and $Dec_{sk}(\{c_0, c_1\}) = m$ as the \textit{Decryption} module - Figure \ref{fig:pcps}. The proposed OT algorithm over ring $R_q$ is as follows.

\vspace{0.05in}
\begin{algorithm}
	\normalfont
	Suppose Alice uses $KeyGen()$ to generate and send a public key to Bob, and keeps the private pairing key to herself. Alice has $l$ $n$-bit binary messages $\{m_1, \cdots, m_l\}$ and $l$ $n$-bit random vectors $\{r_1, \cdots, r_l\}$.
	
	\begin{enumerate}
		\item Alice sends $\{r_1, \cdots, r_l\}$ to Bob. Bob chooses the $c^{th}$ vector $r_c$ in order to acquire $m_c$. Thus Bob generates a random binary vector $K \in R_2^n$ and computes $v$ to send to Alice:
		\vspace{-0.05in}
		\begin{equation}
			v = r_c + Enc_{pk}(K).
		\vspace{-0.05in}
		\end{equation}
		
		\item For all $i \in \{1, 2, \cdots, l\}$, Alice computes the set $\{m'_i\}$ and sends it back to Bob:
		\vspace{-0.05in}
		\begin{equation}
			m'_i = Dec_{sk}(v - r_i) \oplus m_i,
			\vspace{-0.05in}
		\end{equation}
		where $\oplus$ is bitwise XOR. 
		
		\item Bob computes his desired $m_c$ while remaining oblivious to other $m_i$, where $i \neq c$:
		\vspace{-0.1in}
		\begin{equation}
			m_c = m'_c \oplus K.
			\vspace{-0.05in}
		\end{equation}
	\end{enumerate}
\end{algorithm}
 
\vspace{-0.1in}
\subsection{The Zero-Knowledge Proof (ZKP) Primitive}
\vspace{-0.05in}
The ZKP enables an entity to prove to a verifier that it knows a secret value $s$, without revealing any information (including the value of $s$) apart from the fact that it knows the value. 
Similar to the OT primitive, the ZKP primitive is designed using the building blocks the PKC algorithm in section \ref{sec: PKC}. 

\vspace{-0.1in}
\begin{algorithm}
	\normalfont
	Suppose Alice has a secret value $s$ and needs to prove her ownership of $s$ to Bob. 
	
	\begin{enumerate}
		\vspace{-0.05in}
		\item Alice uses $KeyGen(a, s)$ to generate $b$. Alice selects a binary vector $m$, and samples $e', r \leftarrow \mathcal{X}$ to generate $c$: 
		\vspace{-0.05in}
		\begin{equation}
			c = a \cdot r + mt + e', 
			\vspace{-0.05in}
		\end{equation}
		where $t = \lfloor \frac{q}{2} \rfloor$.
		
		Alice sends $\{a, b, m, c\}$ to Bob, and keeps $s$ to herself.  
		
		\item Bob samples $u \leftarrow \mathcal{X}$, and interactively sends it to Alice. 

		\item Alice responds with $x$ to Bob:
		\vspace{-0.05in}
		\begin{equation}
			x = r + s \cdot u.
			\vspace{-0.05in}
		\end{equation}
		
		\item Bob computes and verifies if:
		\vspace{-0.05in}
		\begin{equation} \label{eq: ZKP}
			\lceil (c - a \cdot x + b \cdot u)/t \rfloor \stackrel{?}{=} m,
			\vspace{-0.05in}
		\end{equation}
		where $\lceil \rfloor$ stands for taking the nearest binary integer. 
		
		If the equality of [Eq. \ref{eq: ZKP}] stands, then Alice has successfully proved her ownership of $s$ to Bob.
	\end{enumerate}
\end{algorithm}
\vspace{-0.2in}

\bibliographystyle{ACM-Reference-Format}
\bibliography{paper} 

\end{document}